# Large transverse thermoelectric figure of merit in a Dirac semimetal

[1,2]Junsen Xiang, [1,3]Sile Hu, [1,3]Meng Lyu, [1,3]Wenliang Zhu, [1,3]Chaoyang Ma, [2]Ziyu Chen, [1,4]Frank Steglich, [1,3,5]Genfu Chen, and [1,3,5]Peijie Sun

*1 Beijing National Laboratory for Condensed Matter Physics, Institute of Physics, Chinese Academy of Sciences, Beijing 100190, China*
*2 Department of Physics, Key Laboratory of Micro-Nano Measurement-Manipulation and Physics, Beihang University, Beijing 100191, China*
*3 University of Chinese Academy of Sciences, Beijing 100049, China*
*4 Max Planck Institute for Chemical Physics of Solids, 01187 Dresden, Germany*
*5 Songshan Lake Materials Laboratory, Dongguan, Guangdong 523808, China*

**Thermoelectric (TE) conversion in conducting materials is of eminent importance for providing renewable energy and solid-state cooling. Although traditionally, the Seebeck effect plays a key role for the TE figure of merit $z_S T$, it encounters fundamental constraints hindering its conversion efficiency. Most notably, there are the charge compensation of electrons and holes that diminishes this effect, and the intertwinement of the corresponding electrical and thermal conductivities through the Wiedemann-Franz (WF) law which makes their independent optimization in $z_S T$ impossible [1, 2]. Here, we demonstrate that in the Dirac semimetal $Cd_3As_2$ the Nernst effect, i.e., the transverse counterpart of the Seebeck effect, can generate a large TE figure of merit $z_N T$. At room temperature, $z_N T \approx 0.5$ in a small field of 2 T; it significantly surmounts its longitudinal counterpart $z_S T$ for any field and further increases upon warming. A large Nernst effect is generically expected in topological semimetals, benefiting from both the bipolar transport of compensated electrons and holes and their high mobilities. In this case, heat and charge transport are orthogonal, i.e., not intertwined by the WF law anymore. More importantly, further optimization of $z_N T$ by tuning the Fermi level to the Dirac node can be anticipated due to not only the enhanced bipolar transport, but also the anomalous Nernst effect arising from a pronounced Berry curvature [3 – 6]. A combination of the former topologically trivial and the latter nontrivial advantages promises to open a new avenue towards high-efficient transverse thermoelectricity.**

Thermoelectric (TE) devices can convert heat into electrical energy but may also be used as a heat pump in which electricity can drive a Peltier cooler. Various materials and differing concepts have been employed in order to achieve practical application [1, 2]. Here, the dimensionless figure of merit, $z_S T = T S_{xx}^2 / \rho_{xx} \kappa_{xx}$, plays the key role in characterizing the TE conversion efficiency. It contains the absolute temperature $T$, the Seebeck coefficient $S_{xx}$, the electrical resistivity $\rho_{xx}$ and the thermal conductivity $\kappa_{xx}$. The subscript '$xx$', which denotes diagonal components of the corresponding transport matrix, is intentionally added to distinguish from the off-diagonal counterparts that will define the transverse counterpart $z_N T$. In classical TE materials, optimization of $z_S T$ meets severe limitations. Most important is charge compensation of electrons and holes that contribute oppositely to $S_{xx}$ (Fig. 1**a**). Also, the Lorenz number fundamentally ties $\rho_{xx}$ and the electronic contribution to thermal conductivity $\kappa_{xx}^e$, i.e., $L = \kappa_{xx}^e \rho_{xx}/T$. $L$ coincides with Sommerfeld's constant $L_0 = 2.44 \times 10^{-8}$ W·Ω·K$^2$ for dominating elastic scatterings (the Wiedemann-Franz, WF, law). Though the WF law is frequently violated to some extent at finite temperatures, it makes independent optimization of $\rho_{xx}$ and $\kappa_{xx}$ for high $z_S T$ impossible. Any new designing principle circumventing (some of) these limitations will potentially achieve a big advance in TE application, such as the electron crystal-phonon glass approach, which proposes to minimize $\kappa_{xx}$ while simultaneously keeping $\rho_{xx}$ low by crystal-structure engineering [7].

A number of suited TE materials have recently been revisited and turned out to be topological materials [2, 8]. Except for some basic properties connecting both types of materials [8], e.g., the presence of heavy elements and small band gaps, characteristic reasons for advanced TE transport in topological materials have also been discussed. These include the specific band complexities [9 - 11] and charge relaxation processes [4,12,13] generally related to topological materials. However, the aforementioned topologically trivial limitations for TE performance exist for topological materials as well. Another, largely unexplored, prominent feature of these materials is the strong magnetic-field response, pertinent particularly to topological semimetals [14,15]: Here, in addition to the well-known chiral anomaly in the magneto-electrical conductivity, a strong effect of magnetic fields on the Seebeck effect was recently realized, too. For example, Wang et al. have reported a largely enhanced $z_S T$ from 0.2 to 1.2 by applying a field $B = 7$ T at $T = 370$ K for the Dirac semimetal $Cd_3As_2$ [16]. Moreover, a large, nonsaturating Seebeck effect in quantizing magnetic fields has been theoretically predicted for Dirac/Weyl semimetals [17].

In case a magnetic field $B_z$ is applied orthogonally to the temperature gradient $dT_x$, a transverse counterpart ($S_{xy}$) of $S_{xx}$, known as the Nernst effect (Fig. 1**b**), appears too. Traditionally, $S_{xy}$ has been intensively investigated mainly in elemental Bi and several of its binary alloys, like Bi-Sb [18 – 21], where its potential for TE application was first proposed. The key ingredients for significant values of $S_{xy}(B_z)$ revealed in these works include: A large product of $\omega_c \tau$ ($\omega_c$ is the cyclotron frequency and $\tau$ the relaxation time) and a small Fermi energy $\varepsilon_F$ (see SM for explanation why $S_{xy}$ tracks $\omega_c\tau/\epsilon_F$ in a single-band approximation), as well as a strong electron-hole compensation that gives rise to a significant bipolar effect, see Fig. 1**b**. Interestingly, similar conditions are inherent to a large number of recently discovered topological semimetals [15]. Indeed, a large Nernst power factor, $S_{xy}^2/\rho$, was found in NbP [22] where, however, a large thermal conductivity prevents realization of a sizable transverse figure of merit $z_N T$. Furthermore, an anomalous Nernst effect (ANE) due to a pronounced Berry curvature of the related electronic bands, in additional to the aforementioned topological trivial origins, has been proposed [3 – 6, 23 - 25] for topological conducting materials with Dirac/Weyl nodes sufficiently close to $\epsilon_F$. Experimental verifications of this were reported very recently for NbP [25,26], TaP/TaAs [23] and $Cd_3As_2$ [24].

In this paper, we demonstrate that $Cd_3As_2$ exhibits a sizable transverse figure of merit which surmounts its longitudinal counterpart in a wide temperature range above 100 K, amounting to $z_N T \approx 0.5$ (0.7) at $T = 300$ (350) K in $B_z = 2$ (2.5) T. These $z_N T$ values, after being normalized by the corresponding field, are more than 2 times its longitudinal counter $z_S T$ in this temperature window. More importantly, due to the remove of the two constraints that apply inherently to $S_{xx}$ (see Fig. 1**b**), as well as the theoretically predicted addition of the ANE, the transverse thermoelectricity in topological semimetals appears to be much easier for further optimization by e.g., tuning the Fermi level.

As shown in Fig. 1c, for $Cd_3As_2$ the magnitude of $S_{xy}(T)$ surpasses that of $S_{xx}(T)$ already in weak magnetic fields, e.g. $B_z$ = 1 T, at $T$ > 100 K. This is even clearly demonstrated in Fig. 1d where the ratio $S_{xy}/S_{xx}(T)$ for $B_z$ = 0.5, 1, 2 and 3 T is shown. Furthermore, the initial Nernst coefficient, defined as $N_{ini}$ = d$S_{xy}$ /d$B_z$ at $B_z \to 0$, strongly increases with $T$ and reaches 100 µV/KT at $T \approx$ 250 K (Fig. 1e) before decreasing smoothly. Such $N_{ini}(T)$ profile has been interpreted for topological semimetal NbP [26] as reflecting the $T$-dependent competition between electron- and hole- bands, i.e., the bipolar effect. Subsequently, the existence of an enhanced ANE was also emphasized [25], originating from the shift of the Fermi level $\epsilon_F$ toward the Weyl node with increasing temperature [26]. A larger transverse relative to longitudinal TE response can also be inferred from the so-called thermoelectric mobility $\mu_T = N_{ini}/S_{xx}$ that is as large as its electrical counterpart, the Hall mobility $\mu_H$; the latter surmounts 10 m$^2$ V$^{-1}$s$^{-1}$ at $T$ = 10 K, see Fig. 1f. A large value of $\mu_H$ pushes the strong field limit ($\mu_H B$ > 1) down to below 1 T, where $\omega_c \tau$ > 1 is already realized. These properties, in combination with the extremely low thermal conductivity of $Cd_3As_2$ [27, 28], are mainly responsible for the unprecedented magnitude of transverse thermoelectricity of this compound in low fields.

Figure 2 compiles the full set of magneto-transport coefficients for sample S1 ($\epsilon_F$ = 253 meV, see SM) at selected temperatures in panels **a** – **e**. In panel **f**, we compare the Nernst effect of this sample to the previously reported values in ref. 24. See supplementary Fig. S3 for similar results of sample S4 with a higher $\epsilon_F$ = 273 meV. Significantly, both $S_{xx}(B_z)$ and $S_{xy}(B_z)$ (Fig. 2**a** and 2**b**) are sensitive to small fields. At $T$ > 50 K, they are substantially enhanced up to a certain value of $B_z \leq 3$ T, then the former quantity tends to saturate and the latter assumes a maximum, see Fig. 2**c**, where $S_{xx}(B_z)$ is compared to $S_{xy}(B_z)$ for $T$ = 250 K. There, the value of $S_{xy}(B_z)$ can surpass $S_{xx}(B_z)$ at fields as low as $B_z \approx$ 0.6 T and by 50% at 2 T, as already indicated in the ratio $S_{xy}/S_{xx}(B_z)$ shown in Fig. 1**d**. While $S_{xy}(B_z)$ is substantially larger than $S_{xx}(B_z)$ in moderate fields, we note that the field enhancement to the latter is already sizable and has been reported to cause a marked increase of $z_S T$ in $Cd_3As_2$ [16], as mentioned in the introduction.

As far as TE conversion is concerned, the increase of $\rho_{xx}(B_z)$ (Fig. 2**d**) is almost compensated by the corresponding decrease of $\kappa_{xx}(B_z)$ (Fig. 2**e**), both of which appear in the denominator of $z_S T$. In other words, given that the WF law roughly holds at $T \geq$ 100 K, the thermal conductivity measured in this temperature range is predominantly electronic in origin and $\kappa_{xx}\rho_{xx} \approx \kappa_{xx}^e \rho_{xx} = L_0 T$. Indeed, as demonstrated in Fig. 2**e** for $T$ =150 K, the significant decrease of $\kappa_{xx}(B_z)$ up to $B_z \approx$ 2 T can be explained by the estimated $\kappa_{xx}^e(B_z)$ based on the WF law (dashed line). On the other hand, the phonon contribution to thermal conductivity, $\kappa_{xx}^{ph}(T)$, is negligibly small at $T$ > 100 K compared to $\kappa_{xx}^e(T)$, see Fig. S5. This is due to the intrinsically formed crystallographic vacancies in $Cd_3As_2$, which is Cd-deficient of the ideal $Cd_4As_2$ antifluorite formula [27, 28]. Superior electronic transport properties coexisting with negligible phonon thermal conductivity was also observed in other topological semimetals such as WP$_2$ [29] and ZrTe$_5$ [30]. This situation yields an ideal playground for TE manipulation: Here, $z_S T \approx S_{xx}^2/L_0$ and $S_{xx}$ is the only free parameter; a value of $S_{xx} \geq$ 155 µV/K would lead to $z_S T \geq$ 1. The same argument applies to the Nernst effect as well.

To interpret the large Nernst effect in $Cd_3As_2$, comparison with other semimetals is instructive. Qualitatively similar profiles of both $S_{xx}(B_z)$ and $S_{xy}(B_z)$ as observed in $Cd_3As_2$ have been observed in topological semimetal $Pb_{1-x}Sn_xSe$ with $x$ = 0.23, too [31]. There, low-field values of $S_{xy}(B_z)$ are somewhat lower but still sizable. For example, $S_{xy}(B_z = 2\,T) \approx 50$ µV/K at $T$ = 250 K. These values are, however, much smaller than the corresponding values of $S_{xx}(B_z)$. The situation that $S_{xx}(T, B_z) \gg S_{xy}(T,B_z)$ in a large parameter space indicates a much reduced bipolar transport in the latter compound, qualifying it as a state-of-the-art (longitudinal) TE material [1, 2, 31]. Consequently, rapid increase of both $S_{xx}(B_z)$ and $S_{xy}(B_z)$ in $Pb_{1-x}Sn_xSe$ ($x$ = 0.23) in low fields and at low temperatures ($T$ < 150 K) has been successfully approached by a single, high-mobility band [31]. It imposes a competition between longitudinal and transverse differential logarithmic conductivities $D = \partial \ln\sigma_{xx}/\partial\varepsilon$ and $D_H = \partial\ln\sigma_{xy}/\partial\varepsilon$ at the Fermi level $\varepsilon_F$. It is this competition that determines the characteristic profiles of $S_{xx}(B_z)$ and $S_{xy}(B_z)$. By employing the experimentally obtained conductivity matrix elements $\sigma_{xx}(B_z)$ and $\sigma_{xy}(B_z)$, one can fit the measured $S_{xx}(B_z)$ and $S_{xy}(B_z)$ of $Cd_3As_2$ independently. However, the obtained values of $D_H-D$ from the two fittings are already different at $T$ = 50 K and the discrepancy further increases with $T$ up to room temperature, see supplementary Fig. S6. Similar behavior has been discussed as originating from the increasing electron-hole bipolar effect beyond the one-band hypothesis [31]. This inference applies to $Cd_3As_2$ as well, and is in consistent to the inference made from the $N_{ini}(T)$ profile (Fig. 1**e**), which reveals a maximum near room temperature. Further evidence of bipolar transport in $Cd_3As_2$ comes from the non-linear $\rho_{xy}(B_z)$ behaviors that indicate a dominating electron band and a hole band that increasingly compensates upon warming, see supplementary Figs. S7.

In analogy to $z_ST$ that is defined from $S_{xx}$, its transverse counterpart can be defined from $S_{xy}$ [19],

$$z_N T = T \frac{S_{xy}^2}{\kappa_{xx}\,\rho_{yy}} \,. \qquad (1)$$

Here, $\kappa_{xx}$ and $\rho_{yy}$ take the values measured along $x$ and $y$ directions, respectively. This definition can be easily rationalized for Nernst configuration, where the induced electrical current ($\parallel y$) is orthogonal to the driving thermal current ($\parallel x$), see Figs. 1**b**, 3**a,** and ref. 19. In all our samples of $Cd_3As_2$, $\rho$ is actually rather isotropic at least within the (112) plane that we focus on in this work, see supplementary Fig. S4. We therefore use the simultaneously measured $\rho_{xx}$ (rather than the separately measured $\rho_{yy}$) to compute $z_NT$ in order to reduce systematic errors. As shown in Fig. 3**b**, the superiority of the transverse over the longitudinal TE effect in $Cd_3As_2$ is immediately apparent: $z_NT(B_z)$ exhibits a maximum which, at $T$ > 150 K, is not only higher but also occurs at a lower field compared to the maximum in $z_ST(B_z)$. These maximum values, denoted as $z_NT^{max}$ and $z_ST^{max}$, respectively, are shown as a function of $T$ in Fig. 3**c**. There, $z_NT^{max}$ is increasingly enhanced with temperature over its longitudinal counterpart by, e.g., approximately 50% at $T$ = 350 K. After being normalized by their corresponding fields as $z_NT^{max}/B_z^{max}$, the contrast between the two configurations becomes even more striking, see Fig. 3**d**. Above $T$ = 150 K, the maximum $zT$ value per Tesla of the transverse TE configuration exceeds its longitudinal counterpart by about a factor of two in a wide temperature range.

Despite its marked increase at low fields, the value of $z_NT$ is still inadequate for practical applications. A key parameter yet to be optimized is the Fermi energy $\varepsilon_F$, which determines the magnitude of not only the one-band and bipolar diffusion contributions to $S_{xy}$ [21, 32], but also the ANE [23 - 26]. A reduction of $\varepsilon_F$ is expected to enhance all these contributions in a Dirac semimetal. To clarify this point, we have derived the values of $\varepsilon_F$ for all the employed samples from the experimentally observed quantum oscillations in $\rho_{xx}(B_z)$, see Fig. S8. Note that, within the charge-neutrality condition, $\varepsilon_F$ is measured with respect to the energy of minimal electronic density of states, i.e., the energy of the Dirac node. As shown in Fig. 4, $\varepsilon_F$ spans the range of roughly 250 − 290 meV from sample S1 to S5. Most importantly and instructively, both $S_{xy}^{max}$ and $N_{ini}$, which characterize the magnitude of the Nernst effect, increase monotonically with lowering $\varepsilon_F$, see Fig. 4. A simple extrapolation indicates that $S_{xy}^{max}$ would be greater than 155 µV/K if $\varepsilon_F$ < 200 meV. Unfortunately, our sample syntheses could not yield reliable samples with $\varepsilon_F$ < 250 meV. One batch (A5) reported in ref. [24], with $S_{xy}^{max}$ ≈ 150 µV/K at $B_z$ = 5 T, is reproduced in Fig. 2**f**; the Fermi energy of this sample is estimated to be $\varepsilon_F$ ≈ 205 meV (see SM), which indeed meets our expectation for larger $S_{xy}$, cf. Fig. 4. Given the dominating $\kappa_{xx}^e$ in this temperature range, this very sample should exhibit a maximum $z_NT$ of order unity at $T$ = 200 K. This simple relation between the Nernst effect and $\varepsilon_F$ reflects the aforementioned monotonic enhancement of all quantities contributing to $S_{xy}$ upon decreasing $\varepsilon_F$. By contrast, for the five samples investigated we could not find any significant variation of the zero-field values of $S_{xx}$. As discussed in the introduction, this is most likely due to a competition between bipolar and single-band Seebeck effects, where the former diminishes and the latter enhances the value of $S_{xx}$ upon lowering $\varepsilon_F$ [20].

Our observations of large $S_{xy}(T, B_z)$ values exceeding those of $S_{xx}(T, B_z)$ in a wide parameter space is not specific to $Cd_3As_2$, but is generally expected in topological semimetals. An extraordinarily large value of $S_{xy}$ ($T$ = 109 K; $B_z$ = 9 T) ≈ 800 µV/K was recently observed in NbP, and found to be two orders of magnitude larger than the corresponding $S_{xx}$ value [26]. This is not surprising because in NbP $\varepsilon_F$ = 8.2 meV is very close to the Weyl node, leading to symmetric electron and hole excitations that produce a particularly strong bipolar contribution to $S_{xy}$, as well as an enhanced Berry curvature strength [23, 25] that gives rise to very large value of ANE. Unfortunately, the thermal conductivity of NbP ($\kappa$ ≈ 250 W/Km at $T$ = 300 K) is much too large to obtain a sizable $z_NT$ value [22, 33]. In $WP_2$, too, a large Nernst was observed, e.g., $S_{xy}$ ($T$ = 3.66 K, $B_z$ = 5 T) ≈ 1 mV/K, although at much lower temperatures [34]. As has been discussed, due to the generically high mobility of all the pertinent Dirac/Weyl semimetals, the strong-field limit with $\omega_c\tau$ > 1 can be easily accessed by a field as low as 1 T. There, a larger $S_{xy}$ over $S_{xx}$ is anticipated, resembling the case of dominating transverse electrical conductivity with $\sigma_{xy}$ > $\sigma_{xx}$, see Fig. 1**d** and refs. 26 & 34. This is a field accessible even by commercial permanent magnet like $Nd_2Fe_{14}B$. Recently, Nernst effect in the absence of external field has been observed in a magnetic topological semimetal [35]. This certainly deserves more attention in future.

As already demonstrated for $Cd_3As_2$ [24], a steplike increase of $S_{xy}(B_z)$ at low field arising from enhanced Berry curvature and characterizing the ANE is a significant additive to the conventional (diffusive) Nernst effect. This occurs if the Fermi level is sufficiently reduced towards the Dirac node [6, 23, 25], which may be achieved by a well-controlled sample synthesis. This situation appears to be realized for the case of sample A5 in ref. 24, for which the maximum value of $S_{xy}(B_z)$ is reproduced in Fig. 4. A direct identification of a steplike ANE for our samples seems unlikely, see Figs. 2**b** and S3**b**. However, such a contribution is presumably involved in our samples, too, though being expected to be smaller given their higher $\epsilon_F$. Whether in future one can experimentally sort out the evolution of ANE with $\epsilon_F$ appears to be instructive for improvement of $S_{xy}$. The largely unexplored ANE in this compound, together with the enhanced diffusive contribution of bipolar transport, promises a large room for further optimization of $z_N T$. This is believed to be achievable simply by tuning the Fermi energy (Fig. 4). Apart from their fundamental advantages discussed above, transverse TE devices also have remarkable merits due to their unique configuration geometry. i) The TE conversion output of such devices scales with the sample size along *y* relative to *x*, which allows for various possibilities of shape engineering in order to maximize conversion efficiency [19]; ii) Both electrons and holes of a single material are already involved in the Nernst effect, leaving out the traditional idea of employing both *n* and *p*-type materials in one TE device, see Fig.3**a**; iii) Separation of electrical and heat flows into two orthogonal directions removes the fundamental constraint of the WF law; utilization of anisotropic transport becomes possible through, e.g., combining low $\kappa_{xx}$ and low $\rho_{yy}$ in one material.

**METHODS**

Sample synthesis. Single crystals of $Cd_3As_2$ were grown by self-transport technique. Stoichiometric amounts of cadmium and arsenic were first sealed in an evacuated quartz ampoule. The quartz ampoule was heated to 850℃ and then quenched in liquid nitrogen. The obtained polycrystalline $Cd_3As_2$ was filled in a quartz ampoule again. After evacuating and sealing, the ampoule was placed in a two-zone furnace with a temperature gradient from 575 to 500℃ for 10 days, and then naturally cooled down to room temperature. The single crystals grow in bulk form with well-defined facets, see inset of Fig. S1. X-ray diffraction of powered and single-crystalline samples was performed to confirm the proper crystal structure and the orientation (Fig. S1). The residual resistivity ratio RRR = $\rho_{xx}(300K)/\rho_{xx}(2K)$ is between 5 and 10 for all the samples investigated. This is in agreement with most published works on this compound [16, 27], except for several batches reported in ref. [36].

Transport measurements. The three relevant thermal transport coefficients $S_{xx}$, $S_{xy}$ and $\kappa_{xx}$, as well as resistivity $\rho_{xx}$ were measured simultaneously and adiabatically in a magnetic field $B_z$ applied orthogonal to temperature gradient $dT_x$ (Figs. 1**a**, 1**b** and 1**c** inset). $dT_{xx}$ was monitored by a field-calibrated, thin Chromel-AuFe0.07% thermocouple ($\varphi$ = 25 μm), cf. Fig. S2. All the five samples investigated in this work were bar-shaped by polishing the as-grown samples to typical dimension (3−4) × (1−2) × (0.3−0.5) mm$^3$, see Fig. 1**c** inset. Note that the field-induced Nernst voltage $E_y$ is geometry (more precisely, the length-to-width ratio) dependent. By convention, we adopt the geometry-normalized Nernst effect $S_{xy} = A\, E_y/dT_x$, with $A = L_x/L_y$ being the length ratio of thermal and electrical flow [20]. All the thermal-transport measurements were performed within the (112) plane, with magnetic field $B_z$ applied normal to it. The five samples discussed in this work have slightly different Fermi energies, which were estimated from the respective quantum oscillations in resistivity (Figs. 4 and S8).

**Figures**

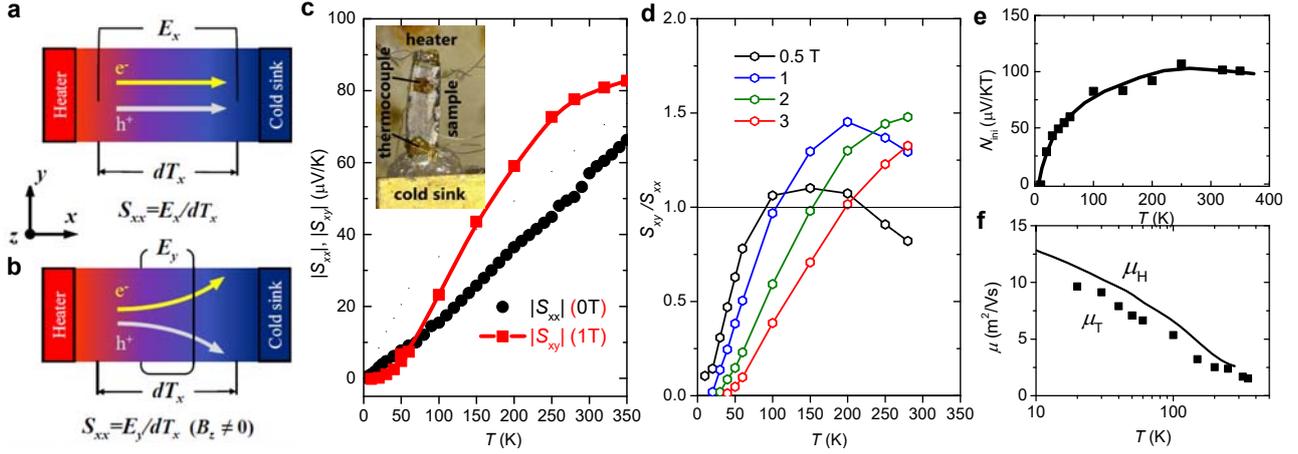

**FIG. 1**: Comparison of longitudinal and transverse thermoelectric effects in $Cd_3As_2$. (**a**, **b**) Schematic illustrations of the Seebeck ($S_{xx}$) and the Nernst ($S_{xy}$) effects for a compensated conductor. They measure the longitudinal and transverse thermoelectric voltages $E_x$ and $E_y$, respectively, over a temperature difference $dT_x$. For the latter effect, an orthogonal field $B_z$ is needed. Note the diminished $S_{xx}$ in (**a**), as compared to the enhanced $S_{xy}$ in (**b**) due to electron-hole bipolar conduction. As heat ($\parallel x$) and electric flux ($\parallel y$) are orthogonal in the latter, constraints of the WF law is circumvented when considering thermoelectric application. (**c**) $S_{xy}(T)$ measured at $B_z = 1$ T, as compared to $S_{xx}(T)$ of $Cd_3As_2$. From the quasi-linear behavior of $S_{xx}(T)$, a rough estimate of the Fermi energy $\varepsilon_F = 305$ meV can be obtained by assuming $S_{xx}(T) = \pi^2/2\, k_B/e\, k_BT/\varepsilon_F$ for simple metals. Inset: The photo image of the sample S1 mounted on sample holder. (**d**) The transverse to longitudinal thermoelectric ratio $S_{xy}/S_{xx}$ at varying field $B_z = 0.5, 1, 2$ and $3$ T is shown as a function of $T$. (**e**) Initial Nernst coefficient $N_{ini}(T)$ defined as $S_{xy}/B_z$ ($B_z \to 0$). (**f**) Thermoelectric mobility $\mu_T$ ($= N_{ini}/S_{xx}$) is compared to its electronic analog, i.e., the Hall mobility $\mu_H$ ($= R_H/\rho_{xx}$). Their values are much larger than 1 $m^2$/Vs (i.e., $\omega_c\tau > 1$ at $B_z = 1$ T) below room temperature, indicating that both electric and thermoelectric responses in small fields are predominantly transverse. Data shown in panels **c**-**f** were measured for sample S1, with **x** $\parallel$ [110] and **z** $\perp$ (112) plane.

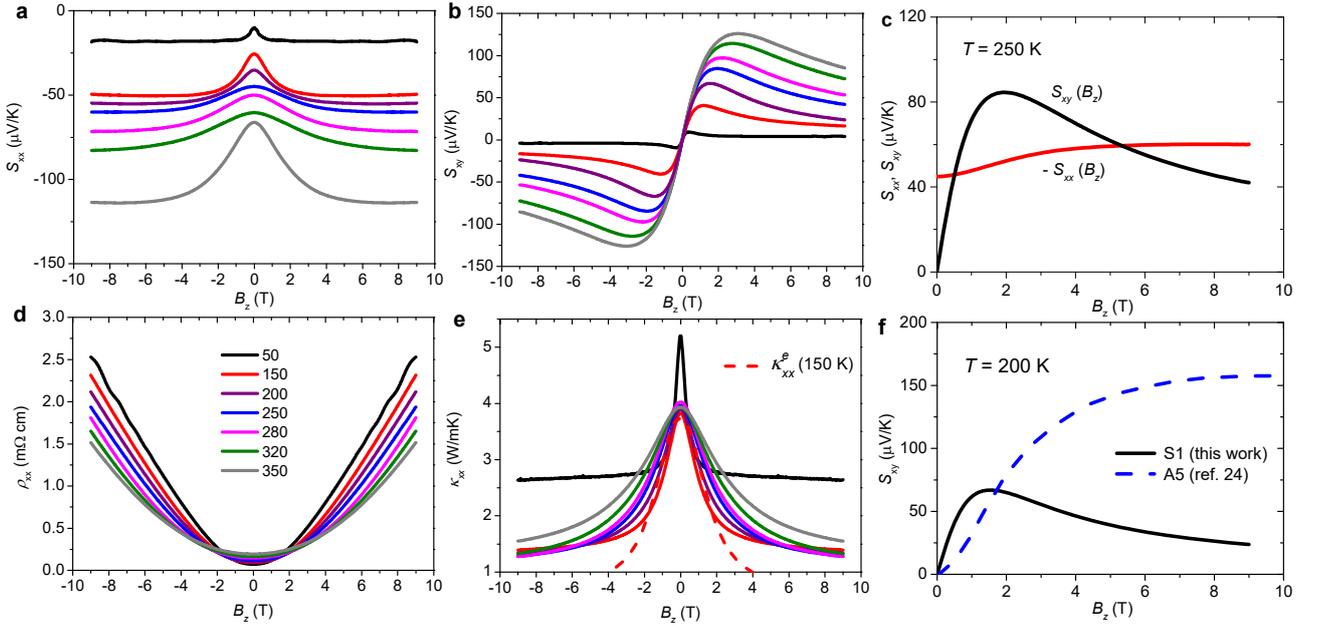

**FIG. 2**: Electrical and thermal transport coefficients recorded in a perpendicular field scan at varying temperatures for sample S1. The Seebeck coefficient $S_{xx}(B_z)$ and Nernst effect $S_{xy}(B_z)$ are shown in (**a**) and (**b**), respectively. In (**c**), $S_{xx}(B_z)$ and $S_{xy}(B_z)$ for $T = 250$ K are compared, with the latter largely exceeding the former between $B_z = 1$ and 4 T. (**d**) and (**e**) show electrical resistivity $\rho_{xx}(B_z)$ and thermal conductivity $\kappa_{xx}(B_z)$ at the corresponding temperatures, respectively. Because the electronic contribution $\kappa_{xx}^e(T)$ dominates $\kappa_{xx}(T)$ at $T > 50$ K (see supplementary Fig. S5), the field enhancement of $\rho_{xx}(B_z)$ and the field reduction of $\kappa_{xx}(B_z)$ are nearly compensated when considering TE application. Namely, $\kappa_{xx}\rho_{xx} \approx L_0 T$ in this temperature window. Indeed, as shown in panel (**e**), the measured $\kappa_{xx}(B_z)$ for $T = 150$ K can be well appraoched by the calculated $\kappa_{xx}^e$ (= $L_0 T/\rho_{xx}$) at $B < 2$ T. This means that the denominator of the $z_S T$ expression, $\rho_{xx}\kappa_{xx}$, approximates to a constant and practically, $S_{xx}$ and $S_{xy}$ become the solely free material parameters that determine the TE performance. (**f**) $S_{xy}(B_z)$ measured at $T = 200$ K for sample S1 is compared to that of A5 reported in ref. 24. The Fermi energy $\epsilon_F$ is 253 and 205 meV for S1 and A5, respectively, see SM.

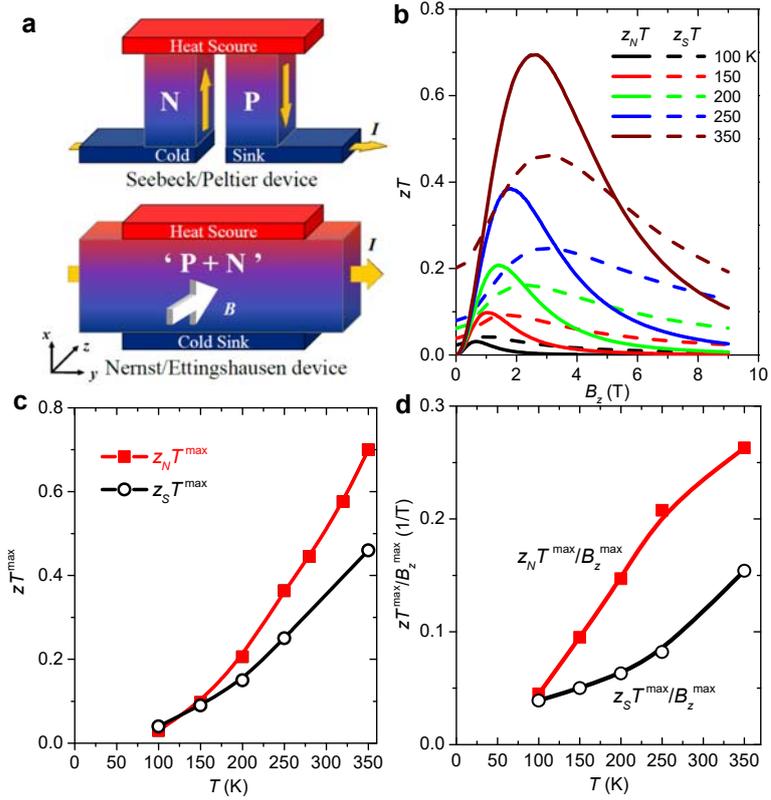

FIG. 3: Enhanced transverse thermoelectricity in $Cd_3As_2$. (**a**) Schematics of longitudinal (upper) and transverse (lower) TE device. While the former needs both p- and n-type materials, the latter uses only one material with, however, inherent contributions from both types of charge carrier through biploar transport. (**b**) $z_NT(B_z)$ in comparison to $z_ST(B_z)$ at selected temperatures for sample S1. (**c**) $z_NT^{max}$ and $z_ST^{max}$, i.e., the maximum values of $z_NT(B_z)$ and $z_ST(B_z)$ observed at certain field $B_z^{max}$, are shown as a function of $T$. (**d**) $zT^{max}$ normalized to the corresponding field $B_z^{max}$ for both longitudinal and transverse configurations, where the latter amounts to more than 2 times the former over a large temperature range above 100 K.

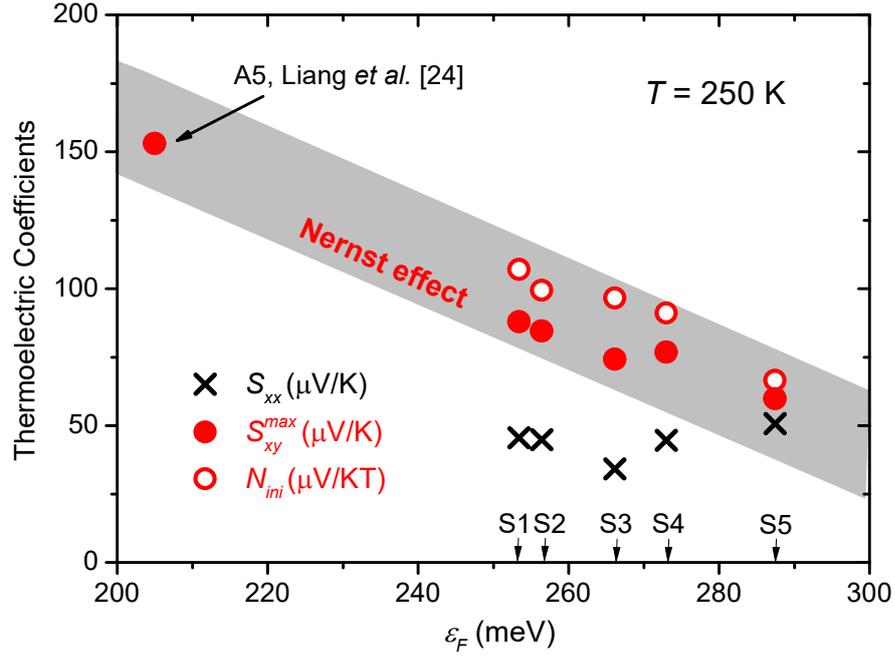

**FIG. 4**: Longitudinal and transverse thermoelectric coefficients as a function of Fermi energy $\varepsilon_F$ for $Cd_3As_2$ at $T = 250$ K. $S_{xy}^{max}$ denotes the maximum values of $S_{xy}(B_z)$, see Figs. 2**b** & S3**b**, $N_{ini}$ the initial Nernst coefficient (cf. Fig. 1**e**). Both quantities increase monotonically with lowering $\varepsilon_F$, owing to the simultansously enhanced single-band and bipolar diffusion contributions. Given the existence of ANE arising from Berry curvature in $Cd_3As_2$, it will also increase with lowering $\varepsilon_F$ [23 - 26]. The data point for sample A5 at $\epsilon_F = 205$ meV (reproduced from ref. 24; see Fig. 2**f**, too) was measured at $T = 200$ K; it presumably would lead to a significant $z_N T$ of unity, see maintext. By contrast, the Seebeck coefficient $S_{xx}$ at zero field reveals no significant $\varepsilon_F$ dependence. This can be easily understood by considering the opposite trend of single-band and bipolar effect in $S_{xx}$ as a function of $\epsilon_F$. While lowering $\epsilon_F$ will enhance the single-band Seebeck effect, it also enhances the bipolar transport which diminishes $S_{xx}$.